\documentclass{aip-cp}
\usepackage{amsfonts,amssymb,amsbsy,amscd}
\usepackage[numbers]{natbib}
\usepackage{rotating}
\usepackage{graphicx}
 \newtheorem{theorem}{Theorem}
 \newtheorem{Definition}{Definition}
  \newtheorem{remark}{Remark}

\def\dfrac#1#2{\displaystyle{#1\over #2}}

\def\bv{{\bf v}}
\def\bV{{\bf V}}
\def\bp{{\bf p}}

\def\Div{\mbox{div}\,}
\def\Rot{\mbox{curl}\,}

\def\bP{{\bf P}}
\def\bB{{\bf B}}

\def\bE{{\bf E}}
\begin{document}

\title{On nonstrictly hyperbolic systems and models of natural sciences reducible to them}

\author[aff1]
{Olga Rozanova}
\eaddress{Corresponding author: rozanova@mech.math.msu.su}
\affil[aff1]{Moscow State University, Moscow, 119991, Russian Federation.}


\maketitle

\begin{abstract} We show that many important models of natural science in their mathematical formulation can be reduced to non-strictly hyperbolic systems of the same kind. This allows the same methods to be applied to them, so that some important results concerning a particular model can be obtained as corollaries of general theorems.
However, in each case, the models have their own characteristics.
The article contains an overview of both methods that are potentially applicable to such cases, as well as known results obtained for a particular model. In addition, we introduce new models and obtain results for them.
\end{abstract}

\section{INTRODUCTION}

This paper is devoted to a special case of inhomogeneous non-strictly hyperbolic systems and related parabolic systems.
We first present general methods that can be applied to these systems. We start with a theorem on the criterion for the existence of a globally smooth solution of the Cauchy problem, then we discuss the existence of a simple and traveling waves, then we describe the stochastic regularization method, which allows one to obtain formulas for representing continuous solutions, and finally, we define a generalized solution for non-strictly hyperbolic systems of a given type, which are written in a non-divergence form.

 Further, we give an example of a non-strictly hyperbolic system of two equations, which corresponds to an oscillatory process. We show that the following models can be reduced to this system and its parabolic counterpart:
\begin{itemize}
\item model of a cold plasma;
\item model of a stratified fluid in the gravity field;
\item model of heat conduction in Reyleigh-B\'enard convection;
\item model of oscillatory flow of blood in arteries;
\item Euler-Poisson equations arising in astrophysics and physics of semi-conductors.
\end{itemize}

Let us denote
$\bV=(V_1, V_2, \dots, V_n)$, $V_i=V_i(t,x)$,  $t\ge 0$, $x\in\mathbb R$, $t\ge 0$, and consider the system
\begin{equation}\label{msh}
\dfrac{\partial \bV }{\partial t} + A(\bV) \dfrac{\partial\bV }{\partial x}=  Q \bV,
\end{equation}
where
$ A(\bV) =V_1 \mathbb E$, $\mathbb E$ is the $n\times n$ unit matrix and
$Q={Q_{ij}}$ is a $n\times n$ constant matrix.
The matrix $A$ has the
 eigenvalues  $\lambda_i(\bV)=V_1$, the respective
eigenvectors $\bv_i= (0,...,\underbrace{1}_{i-th \,place},...,0)$,
$i=1,\dots, n,$ form the basis of the ${\mathbb R}^n$ space.

The  initial conditions are
\begin{equation}\label{cdg}
\bV (0,x)=\bV_0(x)\in C^1(\mathbb R).
\end{equation}

According to the standard classification
(\ref{msh}) is non-strictly hyperbolic (see \cite{Daf}) and has a local in time solution as smooth as initial data.

We consider also a  parabolic counterpart of (\ref{msh})
\begin{equation}\label{msp}
\dfrac{\partial \bV }{\partial t} + V_1\dfrac{\partial\bV }{\partial x}=  Q \bV + B \dfrac{\partial^2\bV }{\partial x^2},
\end{equation}
where
 $B={b_{ij}}$ is a $n\times n$ constant matrix.

 For some choice of $B$ system (\ref{msp}) can be considered as a regularization of (\ref{msh}), which solution allows shocks waves, therefore there arise a problem of justification of convergence of the viscosity solution to a shocks.
 However, we do not touch here this complicated question. In our applications the components of $B$ are fixed constants.

 \section{1. CRITERION OF A SINGULARITY FORMATION}

 The main advantage of (\ref{msh}) is a possibility to study its solution along characteristics $\dot x(t)=V_1$, i.e. to reduce the dynamics to a system of ODEs. Thus, the components of solution obey linear system
 \begin{equation}\label{msV}
\dfrac{d \bV }{d t} =  Q \bV, \quad \dfrac{d x }{d t} = V_1,
\end{equation}
subject to initial data
\begin{equation}\label{msVcd}
\bV(0) = \bV_0 (x_0),\quad x(0)=x_0, \quad x_0\in \mathbb R.
\end{equation}
If we differentiate (\ref{msh}) with respect to $x$ and obtain along characteristics $\dfrac{d x }{d t} = V_1$, $x(0)=x_0$, a quadratically nonlinear system for $\bv=(v_1,...,v_n)$, $v_i=(V_i)_x$, $i=1,...,n$:
\begin{equation}\label{msv}
\dfrac{d \bv }{d t} = -v_1 \bv+
 Q \bv.
\end{equation}

This system can be linearized using Radon's lemma \cite{Radon}, \cite{Riccati}.

\begin{theorem}[The Radon lemma]
\label{T2} A matrix Riccati equation
\begin{equation}
\label{Ric}
 \dot W =M_{21}(t) +M_{22}(t)  W - W M_{11}(t) - W M_{12}(t) W,
\end{equation}
 {\rm (}$W=W(t)$ is a matrix $(n\times m)$, $M_{21}$ is a matrix $(n\times m)$, $M_{22}$ is a matrix  $(m\times m)$, $M_{11}$ is a matrix  $(n\times n)$, $M_{12} $ is a matrix $(m\times n)${\rm )} is equivalent to the homogeneous linear matrix equation
\begin{equation}
\label{Lin}
 \dot Y =M(t) Y, \quad M=\left(\begin{array}{cc}M_{11}
 & M_{12}\\ M_{21}
 & M_{22}
  \end{array}\right),
\end{equation}
 {\rm (}$Y=Y(t)$  is a matrix $(n\times (n+m))$, $M$ is a matrix $((n+m)\times (n+m))$ {\rm )} in the following sense.

Let on some interval ${\mathcal J} \in \mathbb R$ the
matrix-function $\,Y(t)=\left(\begin{array}{c}\mathfrak{Q}(t)\\ \mathfrak{P}(t)
  \end{array}\right)$ {\rm (}$\mathfrak{Q}$  is a matrix $(n\times n)$, $\mathfrak{P}$  is a matrix $(n\times m)${\rm ) } be a solution of (\ref{Lin})
  with the initial data
  \begin{eqnarray*}\nonumber\label{LinID}
  Y(0)=\left(\begin{array}{c}I\\ W_0
  \end{array}\right)
  \end{eqnarray*}
   {\rm (}$ I $ is the identity matrix $(n\times n)$, $W_0$ is a constant matrix $(n\times m)${\rm ) } and  $\det \mathfrak{Q}\ne 0$ on ${\mathcal J}$.
  Then
{\bf $ W(t)=\mathfrak{P}(t) \mathfrak{Q}^{-1}(t)$} is the solution of (\ref{Ric}) with
$W(0)=W_0$ on ${\mathcal J}$.
\end{theorem}

System (\ref{msv}) can be written as (\ref{Ric}) with
\begin{eqnarray*}\label{M}
W=\bv,\quad
M_{11}=(0),\quad
 M_{12}=(1, 0, ..., 0), \quad
M_{21}=(0, 0, ..., 0)^T,\quad
M_{22}=Q.\\\nonumber
\end{eqnarray*}
Thus, we obtain the linear Cauchy problem
\begin{equation}\label{linv}
\left(\begin{array}{c}
\dot q\\
\dot u_1
\\
\dots
\\
\dot u_n
\end{array}
\right) = \left(
\begin{array}{ccccc}
0 & 1 & 0 &\ldots & 0\\
0& Q_{11} & Q_{12} & \ldots & Q_{1n}\\
\vdots & \vdots & \ddots & \vdots\\
0&Q_{n1} & Q_{n2} & \ldots & Q_{nn}
\end{array}
\right)\,\left(\begin{array}{c}
q\\
u_1
\\
\dots
\\
 u_n
\end{array}
\right),\quad \left(\begin{array}{c}
q(0)\\
u_1(0)
\\
\dots
\\
u_n(0)
\end{array}
\right) =\left(\begin{array}{c}
1\\
v_1(0)
\\
\dots
\\
v_n(0)
\end{array}
\right),
\end{equation}
where $v_i(0)= (V_{0i})_x (x_0)$, $x_0\in \mathbb R$, $i=1,...,n$.

Let us take into account that the singularity formation for the hyperbolic systems means a finite time blow up either solution itself or its first derivatives, then we can conclude that the solution itself which is described by linear system (\ref{msV}) with data (\ref{msVcd}), do not blows up. Further, it follows from Radon's theorem that $v_i(t)=\frac{u_i(t)}{q(t)}$. Therefore
 the derivatives blow up at the moment, when $q(t)$, a part of solution to (\ref{linv}), vanishes.

 We summarize this result as follows.
\begin{theorem}\label{T1}

\begin{enumerate}
  \item The behavior of the solution of the Cauchy problem (\ref{msh}), (\ref{cdg}) along the characteristics can be completely described by linear dynamics with the help of the systems  (\ref{msV}) and  (\ref{linv}).
  \item If for every point $x_0\in \mathbb R$  the scalar function $q(t)$, the component of solution  to (\ref{linv}), does not vanish for all $t>0$, then the solution to  (\ref{msh}), (\ref{cdg}) keeps classical smoothness for all $t>0$.
  \item  Otherwise, if there exists a point $x_0\in \mathbb R$ such that the solution  to (\ref{linv}) is such that $q(t_*)=0$ for a finite $t_*>0$, then the solution to  (\ref{msh}), (\ref{cdg}) blows up is a finite time. The exact time $T_*$ of the blow up is
\begin{eqnarray*}\label{T*}
  T_*= \min\limits_{x_0\in \mathbb R} \left(t_*>0\, \Big|\, q(t_*)=0 \right).
  \end{eqnarray*}
\end{enumerate}
\end{theorem}

\begin{remark}
Although $q(t)$ can always be found explicitly, it is a quasi-polynomial, the solution to the transcendental equation $q(t)=0$ for $n>2$ can usually  be found only numerically.
\end{remark}

\section{2. SIMPLE WAVES}
\begin{Definition}
A solution of (\ref{msh}) or (\ref{msp}) is called a simple wave if there exists a functional dependence
 $V_i=V_i(U(t,x))$, $i=2,...,n$, for some sufficiently smooth function $U(t,x)$.
\end{Definition}

For example, we can consider a simple wave of the form $V_i=V_i(V_1)$, $i=2,...,n$.
A particular important class of simple waves is the class of traveling waves, when $U(t,x)=\xi=x-w t$, $w=\rm const$.

If we substitute  $V_i=V_i(V_1)$ to (\ref{msh}), we get
\begin{equation}
\label{mshsv}
\begin{array}{l}
\dfrac{d V_i}{d V_1} = \frac{Q_{i1}V_1+\sum\limits_{j=2}^n Q_{ij}V_j}{Q_{11}V_1+\sum\limits_{j=2}^n Q_{1j}V_j}, \quad i=2,...,n.
\end{array}
\end{equation}

Traveling waves   $V_i=V_i(\xi)$ for (\ref{msh}) satisfy the system
\begin{eqnarray*}
\label{mshtv}
\begin{array}{l}
\dfrac{d V_i}{d \xi} = \frac{\sum\limits_{j=1}^n Q_{ij}V_j}{V_1-w}, \quad i=1,2,...,n.
\end{array}
\end{eqnarray*}

We can see that for $n=2$ the relation between $V_1$ and $V_2$ can be found explicitly in elementary functions, since the system reduced to one fractional linear equation. For some matrices $Q$ the number of equations in (\ref{mshsv}) reduces, it signifies that  the first integrals exists and $V_i=V_i(V_1,..,V_j),$ $1<j<n$, $i=j+1,...,n$. If $j=n-1$, then there is a full set of first integrals.

\begin{remark}
 The examples show that properties of simple waves can drastically differ from the properties of  solutions of general form.
\end{remark}

The construction of simple waves for a parabolic system (\ref{msp}) is more complicated, since it reduces to higher order ODE systems, usually non-integrable. However, as we will show by examples, there are happy exceptions.

\subsection{3. STOCHASTIC REGULARIZATION}

The method of stochastic regularization was applied first to the Hopf equation, including multidimensional case, in \cite{AKR2013}, and then it turned out that the scope of its application is sufficiently wider. Namely, it was used for the regularization of the solution of the scalar  a scalar conservation laws \cite{AR2013},
 for a system in Riemann invariants \cite{R2014},
for  two-dimensional transport equations on a rotating plane \cite{RU2021}. The main idea is to consider a stochastic perturbation of intensity $\sigma>0$ along characteristics and then introduce a joint density of distribution of the particle position and "velocity" (the solution of initial equation not looking at its physical sense), and consider new system of PDEs describing averaged density and averaged "velocity". This system is integro-differential and its solution tends as $\sigma \to 0$ to the solution of the initial equation provided it is continuous, and to the solution in the sense of "free particles", otherwise. Thus, the method of stochastic regularization is very convenient instrument to construct rarefaction waves, but the limit shock wave is different from  the standard entropy shock.

For  system (\ref{msh}), the stochastic regularization method can be applied, and its justification does not differ from the case of the Hopf equation \cite{AKR2013}, although it is associated with technical difficulties.

\medskip

As a first stem we consider a stochastic differential system which describes a stochastic perturbation of the trajectory $x(t)$, such that $\dot x(t)=V_1(t, x(t))$, $x(0)=x_0$:
\begin{eqnarray*}
d X(t)&=& \mathfrak V_1(t) dt + \sigma dW,\\
d \mathfrak V (t)& =& Q \mathfrak V (t),
\end{eqnarray*}
where $W$ is a standard Wiener process, $\sigma>0$, $\mathfrak V (t)=\bV (t, X(t))$, $\mathfrak V_1 (t)=V_1 (t, X(t))$. Here $X(t)$ and $\mathfrak V=(\mathfrak V_1,..., \mathfrak V_n)$ are random processes starting from the initial point $x_0, \bV_0(x_0)$.

Let $P(t,x, v_1,..,v_n)$ be the joint probability density function for the processes $(X,\mathfrak V (t))$ satisfying the Cauchy problem for the Fokker-Planck equation,
\begin{eqnarray*}
\frac{\partial P}{\partial t }\,+\,v_1 \frac{\partial P
}{\partial x } \, &+& \, \sum \limits_{i=1}^n\,\frac{\partial
}{\partial v_i } \, \left(\sum \limits_{j=1}^n Q_{ij} v_j\,P\right)\, +
\frac{1}{2}\, \sigma^2 \,\frac{\partial^2 P}{\partial
x^2}=0,
\\
  P_0(t,x, v_1, \dots ,v_n)&=&\prod\limits_{i=1}^n \delta(v_i-V_i(0,x))\, f_0(x), \quad f_0\in L_1(\mathbb R).
\end{eqnarray*}

We introduce the following functions
\begin{eqnarray*}
\rho(t,x) &=&
\int\limits_{\mathbb R} P(t,x,v_1, \dots ,v_n)\, dv_1 \dots dv_n,
\\
\hat\bV(t,x)&=& \frac {1}{\rho(t,x)}\int\limits_{\mathbb R} v P(t,x,v_1, \dots ,v_n)\, dv_1\dots dv_n,\quad v=(v_1,...,v_n).
\end{eqnarray*}
After computations we obtain the system
 \begin{eqnarray*}
 \dfrac{\partial \rho  }{\partial t} &+&   \dfrac{\partial \rho \hat V_1 }{\partial x}=  \frac{\sigma^2}{2}
\dfrac{\partial^2 \rho  }{\partial x^2},\\
 \dfrac{\partial \rho \hat\bV }{\partial t} &+&   \dfrac{\partial \rho \hat V_1 \hat\bV }{\partial x}=  Q \rho \hat\bV+ \frac{\sigma^2}{2}
\dfrac{\partial^2 \rho  \hat\bV }{\partial x^2}-\int\limits_{\mathbb R} \, (v-\hat \bV) (v_1-\hat V_1) P_x \, d v.
\end{eqnarray*}
Similarly to \cite{AKR2013}, one can prove that
$
\hat\bV \to \bV,$ $\sigma\to 0,$ $t>0, $ $ x\in\mathbb R,$
for a {\it continuous} $\bV$.

\section{4. GENERALIZED SOLUTION TO THE CAUCHY PROBLEM}

If we still want to consider the solution to the Cauchy problem   (\ref{msh}), (\ref{cdg}) after the moment of blowup, we encounter sufficient difficulties. The problem is that system (\ref{msh})
is not written in a {{\it divergence}} form, so we cannot transfer derivatives to test functions, as is always done when determining the generalized solution of conservation laws. We use a trick, hinted at in the previous section, where the density function naturally arose. We will have to supplement  system (\ref{msh}) with a new equation, after which we can already write it in a conservative form. The solution of (\ref{msh})
turns out to be part of the solution of the new augmented system. Note that the new system is no longer hyperbolic, which leads to the appearance of a delta function in the velocity component, just as it was for the system of gas dynamics "without pressure" \cite{Shelk2009}.

As usual, we start with smooth solutions and complete the system with the continuity equation:
\begin{eqnarray*}
\dfrac{\partial \rho }{\partial t} &+&  \dfrac{\partial \rho V_1}{\partial x}=0,\\
\dfrac{\partial \rho \bV }{\partial t} &+&   \dfrac{\partial  \rho V_1\bV }{\partial x}=  \rho Q \bV.
\end{eqnarray*}
This system has a divergent form, and we can define the usual generalized solution in the sense of an integral identity.

\medskip

In fact, we modify to our case the definition of strong singular  solution according to V.M. Shelkovich \cite{Shelk2009}.
Let
$\Gamma$ be a $C^1$ be a smooth curve given as $S(t,x)=0$, $S_x\ne 0$, dividing the $(t,x)$ -- plane into two part, such that
$\mathbb R^2=\Omega_-\bigcup\Gamma\bigcup \Omega_+ $. Further, let
$\delta(\Gamma)$ be the $\delta$ - function, concentrated on $\Gamma$,
 $e(t,x)\Big|_{\Gamma}$ be its amplitude,
 $v_\delta(t,x)= -{S_t}/{S_x}$ be its velocity, $\mathbb I=v_\delta (1,...,1)^T$,
$\frac{\delta \phi(t,x)}{\delta t}\Big|_{\Gamma}=\left(\dfrac{\partial \phi }{\partial t}+v_\delta \dfrac{\partial \phi }{\partial t}\right)\Big|_{S(t,x)=0}$, \cite{Kanwal}.

\begin{Definition}
We call a vector-function $\bV\in C^1(\Omega_-)\cap C(\bar \Omega_-)\cap C^1(\Omega_+)\cap C(\bar \Omega_+) $
the generalized solution of system (\ref{msh}),
if for every test function $\phi(t,x) \in \mathcal D (\mathbb R^2)$ and every couple of functions $(\hat \rho(t,x), e(t,x))$, $\hat \rho\in  C^1(\Omega_-)\cap C(\bar \Omega_-)\cap C^1(\Omega_+)\cap C(\bar \Omega_+)$, $e(t,x)\in C(\mathbb R^2)$,
such that
\begin{eqnarray*}
\label{base1}
\begin{array}{l}
P(t)=\int\limits_{\mathbb R} \hat \rho (t,x) \, dx + e(t,x)\Big|_{\Gamma} = \rm const,
\end{array}
\end{eqnarray*}
the identity
\begin{eqnarray*}
\label{base1}
\begin{array}{l}
\int\limits_{\mathbb R^2} \hat \rho \left[\bV (\phi_t+ V_1 \phi_x) - Q \bV\phi \right]\, dx \, dt + \int\limits_{\Gamma} e(t,x) \mathbb I \frac{\delta \phi}{\delta t}
\frac{dl}{\sqrt{1+v^2_\delta }}=0
\end{array}
\end{eqnarray*}
holds.
\end{Definition}

As always, when defining a generalized solution, the question of its uniqueness arises. Traditionally, to single out a unique generalized solution, they use
the geometric entropy condition 
\begin{eqnarray*}
\label{accept}
 \min\{{V_1}_-, {V_1}_+\} \le  \dot \Phi(t) \le \max\{{V_1}_-, {V_1}_+\},
\end{eqnarray*}
where $x=\Phi(t)$ is a result of resolving the inexplicit relation $S(t,x)=0$, ${V_1}_\pm$ is the limit values of $V_1(t, x)$ as $x\to \Phi(t)\pm 0.$
It implies that characteristics from both sides of the shock come to the shock.

Solving the Riemann problem for a specific system (\ref{msh})  is a very difficult problem even for the case $n=2$.
We will not touch it here and only note that non-uniqueness can arise here not only for the shock wave, but also for the rarefaction wave. This phenomenon is connected precisely with the non-strict hyperbolicity.

\section{5. MAIN EXAMPLE}

Now we restrict ourself by a particular case of system (\ref{msh}), and are going to show that many important models of physics can be reduced to it or its parabolic counterparts. In this example
$n=2$, $\bV=(V,U)$,
$Q= \left(\begin{array}{cc}0 & -1\\ 1& 0 \end{array}\right)$, such that the system takes the form
\begin{eqnarray}\label{me}
\dfrac{\partial V }{\partial t}  + V \dfrac{\partial V}{\partial x}= - U,\quad
\dfrac{\partial U }{\partial t} +  V \dfrac{\partial U}{\partial x} = V.
\end{eqnarray}
Along characteristics given by equation $\dot x(t)=V(t,x(t)$ we have
\begin{eqnarray*}
\dot V = -U, \qquad \dot U = V,
\end{eqnarray*}
therefore
\begin{eqnarray*}
\dfrac {d (V^2+U^2)}{dt }= \rm const.
\end{eqnarray*}
Thus, the solution is bounded provided it is smooth and
\begin{eqnarray*}
\max\limits_{t>0, \, x\in\mathbb R} {\sqrt{V^2+U^2}}= \max\limits_{x\in\mathbb R} {\sqrt{V_0^2+U_0^2}}.
\end{eqnarray*}

\subsection{5.1. Singularities formation}
The dynamics along characteristics of  derivatives
$(V_x,U_x)=(v,u)$ is as follows:
 \begin{equation}\label{sq}
  \dot v= -u-v^2, \qquad \dot u = v (1 -u).
 \end{equation}
On the phase plane $u,v$ we have one equilibrium point $ (0,0) $, a center. System (\ref{sq}) has
 the first integral
\begin{eqnarray*}
v^2+2 u -1 =C (u-1)^2,
\end{eqnarray*}
corresponding to a second-order curve. The type of this curve depends on the sign of
$
D=v^2+2 e -1.
$
If $ D <0 $, then the phase curve is an ellipse, the derivatives for $t> 0 $, the period is $2\pi$.
Otherwise, the phase curve is a parabola ($ D = 0 $) or a hyperbola ($ D> 0 $), the derivatives  become infinite in a finite time.

This results can be summarised as a theorem \cite{RChZAMP21}.
\begin{theorem}\label{T3}   For the existence and uniqueness of continuously differentiable  $ 2 \pi- $ periodic in time  solution $ (V,U)$ of
\begin{eqnarray*}
\dfrac{\partial V }{\partial t}  + V \dfrac{\partial V}{\partial x}&=& - U,\quad
\dfrac{\partial U }{\partial t} +  V \dfrac{\partial U}{\partial x} = V, \\ (V,U)|_{t=0}&=&(V_0,U_0)\in C^1({\mathbb R})
\end{eqnarray*}
 it is necessary and sufficient  that  inequality
 \begin{eqnarray*}\label{cond}
 (V'_0(x))^2+2 U'_0(x)-1<0
 \end{eqnarray*}
holds at each point $ x \in  \mathbb R$.

If there exists at least one point $ x_0 $ for which the opposite inequality  holds, then the derivatives of the solution become infinite in a finite time.
\end{theorem}

\begin{remark} In this simple case we obtained the criterion of singularity formation independently, but it also follows from Theorem \ref{T1} as a corollary.
\end{remark}

\subsection{5.2. Traveling waves for $B\ne 0$}

To find traveling waves we have to substitute $\bV=\mathbb V(\xi)$, $\xi=x-wt$, to (\ref{msp}).

The resulting ODE system is
\begin{eqnarray*}
(-w {\mathbb E}+ A({\mathbb V})){\mathbb V}'=  Q {\mathbb V} + B {\mathbb V}''
\end{eqnarray*}
or
\begin{eqnarray*}
(-w + {\mathbb V}_1){\mathbb V}'=  Q {\mathbb V} + B {\mathbb V}''.
\end{eqnarray*}
It our particular 2D case ${\mathbb V} =({\mathcal V, \mathcal U})$, therefore
\begin{eqnarray*}\label{q1}
(-w + {\mathcal V}){\mathcal V}'&=& - {\mathcal U} + b_{11} {\mathcal V}'' +b_{12} {\mathcal U}'',\\
(-w + {\mathcal V}){\mathcal U}'&=&  {\mathcal V} + b_{21} {\mathcal V}'' +b_{22} {\mathcal U}''.
\end{eqnarray*}

 As we noticed in Sec.2, for $B=0$ the system can always be integrated, the integral found in  \cite{RChZAMP21}. For $B\ne 0$ it is not integrable in the general case.

\section{6. A ``COLD'' (ELECTRON) PLASMA}

The equations of electron plasma consist of hydrodynamic equations together with  Maxwell's equations \cite{GR75}, \cite{ABR78}:
\begin{eqnarray*}
\label{base1}
\begin{array}{l}
\dfrac{\partial n }{\partial t} + \Div(n \bV)=0\,,\quad
\dfrac{\partial \bP }{\partial t} + \left( \bV \cdot \nabla \right) \bp
= e \, \left( \bE + \dfrac{1}{c} \left[\bV \times  \bB\right]\right) - q\bP  + \nu \Delta \bP,\vspace{0.5em}\\
\gamma = \sqrt{ 1+ \dfrac{|\bP|^2}{m^2c^2} }\,,\quad
\bV = \dfrac{\bP}{m \gamma}\,,\vspace{0.5em}\\
\dfrac1{c} \frac{\partial \bE }{\partial t} = - \dfrac{4 \pi}{c} e n \bV
 + {\Rot}\, \bB\,,\quad
\dfrac1{c} \frac{\partial \bB }{\partial t}  =
 - {\Rot}\, \bE\,, \quad \Div \bB=0\,,
\end{array}
\end{eqnarray*}
where
$e, m$ is the charge and mass of the electron ( $e < 0$),
$c$  is the speed of light,
$ n, \bP, \bV$ are density, momentum and speed of electrons,
$\gamma$ is the Lorentz factor,
$ \bE, \bB $ are vectors of electric and magnetic fields. It this model it is assumed that ions are immobile, $q>0$ is a coefficient of intensity of electron-ion interactions.

In dimensionless quantities under the following assumptions:
\begin{itemize}
 \item  $\gamma=1$, $\bP=\bV$ (non-relativistic case);
\item  $\bB=0$ (electrostaticity);
\item $\nu=0$ (non-viscous case);
\item  $q=0$ (no interaction with ions);
\item  $\bV=(V(t,x),0,0)$, $\bE=(E(t,x),0,0)$;
\end{itemize}
we obtain the following system \cite{CH18}
\begin{equation}\label{q1}
\dfrac{\partial V }{\partial t}  + V \dfrac{\partial V}{\partial x}=- E,\qquad
\dfrac{\partial E }{\partial t} + V \dfrac{\partial E}{\partial x} = V,
\end{equation}
where in the notation of Sec.5
$(V,U)=(V,E)$. It coincides with (\ref{me}).

Thus, as
a corollary of Theorem \ref{T1} we obtain
a criterion for the singularity formation.
The singularity means the ``gradient catastrophe'' for $V$ and $E$ and the delta-singularity for the density.

If we do not neglect viscosity, we cannot apply the previous methods of study of singularity formation. Moreover, we can expect that the presence of viscous term would eliminate the singularity at all. It is a complicated issue, but, as follows from our consideration below,  this hypothesis does not seem to be correct. Indeed, the first step for the justification of the viscosity method for the shock solution of the Riemann problem is the construction of a traveling wave
$(V,E)=({\mathcal V}(\xi), {\mathcal E}(\xi))$, $ \xi= x-wt $,
coinciding the left and right states,  such that ${\mathcal V}'(\pm\infty)= {\mathcal E}'(\pm\infty)=0$. However, the traveling wave does not exist on the whole axis $\xi\in\mathbb R$, and this is the main obstacle. Moreover, the presence of viscosity does not simplify the structure of the traveling wave.

Thus, system (\ref{q1}) for the traveling wave takes the form
\begin{eqnarray*}\label{q1}
(-w + {\mathcal V})\,{\mathcal V}'= - {\mathcal E} + \nu {\mathcal V}'',\qquad
(-w + {\mathcal V})\,{\mathcal E}'=  {\mathcal V},
\end{eqnarray*}
which can be rewritten as one equation
\begin{eqnarray}\label{Vnu}
\nu \,{\mathcal V}'''- ({\mathcal V}-w)\,{\mathcal V}'' - ({\mathcal V}')^2- \frac{{\mathcal V}}{{\mathcal V}-w} =0.
\end{eqnarray}
 Its solution cannot be found explicitly, however we can study small perturbations of the zero steady state.
To this aim we assume
${\mathcal V}=\varepsilon {v}+o(\varepsilon)$, $\varepsilon<<1$, which implies the linear equation
\begin{eqnarray*}\label{q1}
\nu w\,v'''+ w^2\,v'' + v =0.
\end{eqnarray*}
Roots of the characteristic equation are the following:
\begin{itemize}
\item
$\nu=0$: $\lambda_{1,2}=\pm \,  \i\,  w  $, the traveling wave is periodic with respect to $\xi$;

\item $\nu>0$: $\lambda_{1,2}=\alpha\pm \,  \i\,  \beta $, $\lambda_3\in \mathbb R$, the traveling wave is not periodic.
\end{itemize}

Now we present the picture, obtained numerically, which show a drastic difference between viscous and non-viscous traveling waves for solutions of (\ref{Vnu}). The computations are made by the Runge-Kutta method in the MAPLE packet for
$w=2$, ${\mathcal V}(0)=1$, ${\mathcal V}'(0)=0$.

\begin{figure}[h!]
\begin{minipage}{0.4\columnwidth}
\centerline{
\includegraphics[width=0.7\columnwidth]{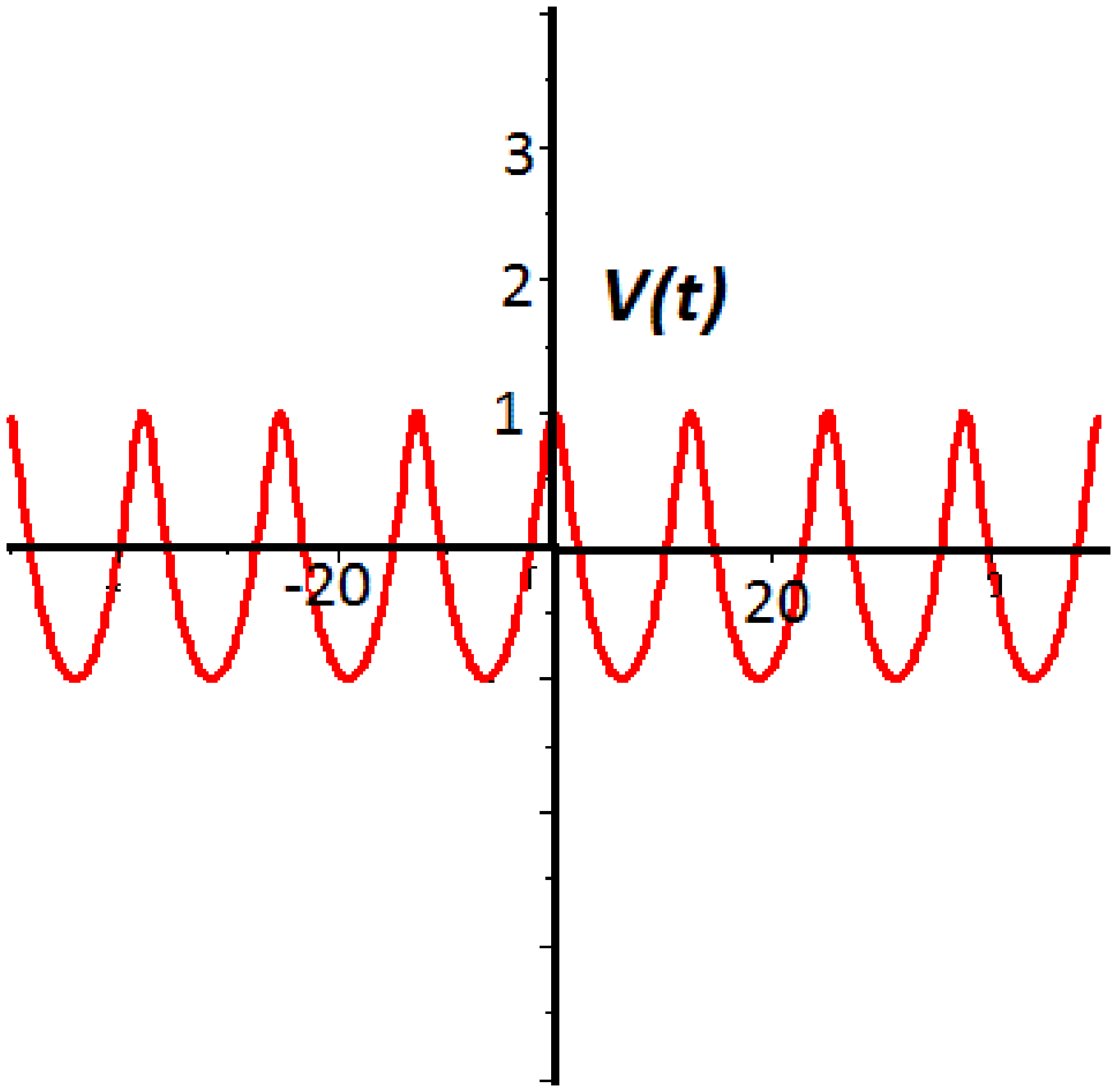}}
\end{minipage}

\begin{minipage}{0.4\columnwidth}
\centerline{
\includegraphics[width=0.7\columnwidth]{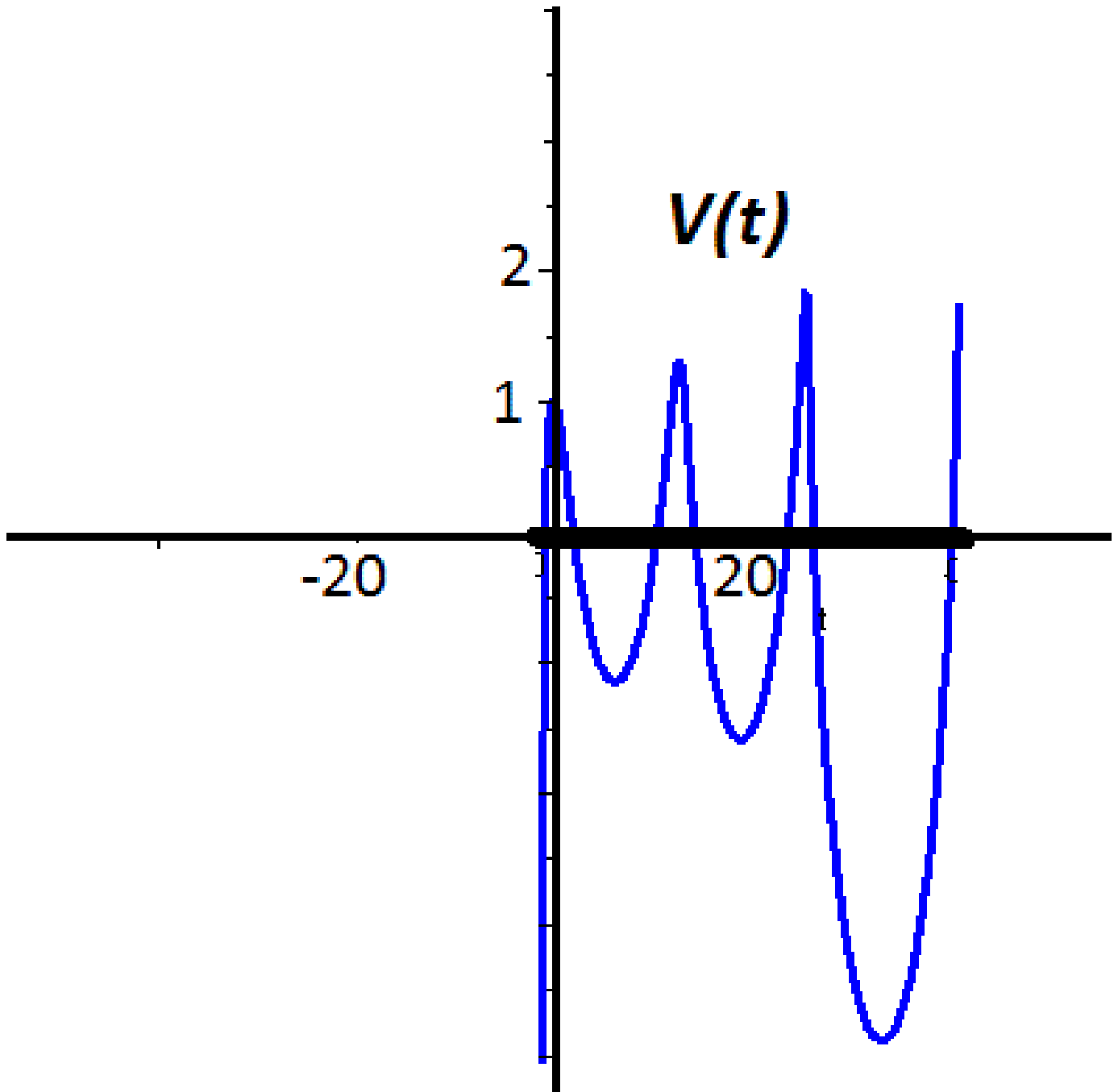}
}
\end{minipage}
\caption
{Periodic traveling wave for $\nu=0$ (left). Traveling wave for $\nu=0.2$,
the solution exists on a bounded interval (right).}
\end{figure}

\section{7. MODEL OF PERTURBATION OF THE BASIC STATE OF REST FOR  HEAT CONDUCTION IN REYLEIGH-B\'ENARD CONVECTION  AND FOR STRATIFIED FLUID  IN THE GRAVITY FIELD }

1. The model of perturbation of the basic state of rest with heat conduction in Reyleigh-B\'enard convection was introduced by P. Drazin, 2002 \cite{Drazin2002}.

In dimensionless variables it has the form
\begin{eqnarray}\label{RB}
\dfrac{\partial V }{\partial t}  + V \dfrac{\partial V}{\partial x}=- \Theta + \nu \dfrac{\partial^2 V}{\partial x^2},\qquad
\dfrac{\partial \Theta }{\partial t} + V \dfrac{\partial \Theta}{\partial x} = V + \kappa \dfrac{\partial^2 \Theta}{\partial x^2},
\end{eqnarray}
where  $x$ is the vertical coordinate,
 $V$ and $\Theta$ are  perturbations of the vertical component of velocity and temperature,
 $\nu$ and $\kappa$ are coefficients of kinematic viscosity and of the heat diffusion.

2. The model of a stratified fluid near a state of rest in a gravitational field was considered by V.G. Baidulov in  2014 \cite{Baidulov}.

It has the form
\begin{eqnarray*}\label{q1}
\dfrac{\partial V }{\partial t}  + V \dfrac{\partial V}{\partial x}=- gS + \bar\nu \dfrac{\partial^2 V}{\partial x^2},\qquad
\dfrac{\partial S }{\partial t} + V \dfrac{\partial S}{\partial x} =\frac{1}{\Lambda} V + \bar\kappa \dfrac{\partial^2 S}{\partial x^2},
\end{eqnarray*}
where $x$ is the vertical coordinate,
 $V$ and $S$ are  perturbations of the vertical component of velocity and salinity,
$g$ is the acceleration of gravity, $\Lambda$ is the stratification scale ($N=\sqrt{g/\Lambda}$ is the buoyancy frequency),
 $\nu$ and $\kappa$ are coefficients of kinematic viscosity and of the diffusion of salt.

In dimensionless variables the system takes the form
\begin{equation}\label{SF}
\dfrac{\partial V }{\partial t}  + V \dfrac{\partial V}{\partial x}=- S +\nu \dfrac{\partial^2 V}{\partial x^2},\qquad
\dfrac{\partial S }{\partial t} + V \dfrac{\partial S}{\partial x} = V + \kappa \dfrac{\partial^2 S}{\partial x^2},
\end{equation}
where $\nu= {\bar \nu N}/{\bar V^2}$ and $\kappa = {\bar \kappa N}/{\bar V^2}$ are analogous to the Reynolds number and P\'eclet number.

We see that systems (\ref{RB}) and (\ref{SF}) coincide up to notation. Further, $(V,U)=(V,\Theta)$ for the first case and $(V,U)=(V,S)$ for the second case.

Thus, in the non-dissipative  limit $\nu=\kappa=0$ both systems  coincide with  (\ref{me}), and therefore we get a criterion for the singularity formation as a corollary of Theorem \ref{T3}.

Let us study traveling waves in the dissipative case, for example, for the 1D stratified fluid. The system for  traveling waves is
\begin{eqnarray*}\label{q1}
(-w + {\mathcal V})\,{\mathcal V}'= - {\mathcal S} + \nu {\mathcal V}'',\qquad
(-w + {\mathcal V})\,{\mathcal S}'=  {\mathcal V}+ \kappa {\mathcal S}'',
\end{eqnarray*}
it can be written as one nonlinear ODE of the 4th order for $\mathcal S$, it is non-integrable.

Small perturbations of the zero steady state under the assumption
${\mathcal S}=\varepsilon {s}+o(\varepsilon),\quad \varepsilon<<1,$
satisfy the following linear equation
\begin{eqnarray*}\label{q1}
\nu \kappa \,s''''+
(\nu+\kappa) w\,s'''+ w^2\,s'' + s =0.
\end{eqnarray*}
Roots of the characteristic equation are the following:
\begin{itemize}
\item
for $\nu=\kappa=0$: $\lambda_{1,2}=\pm \,  \i\,  w  $, the traveling wave is periodic with respect to $\xi$;

\item for $\nu>0$, $\kappa>0$: $\lambda_{1,2}=\alpha\pm \,  \i\,  \beta $,

$\lambda_{3,4}\in \mathbb R$, the same sign, for large $w$,

or $\lambda_{3,4}=\gamma\pm \,  \i\,  \delta$, for small $w$,

the traveling wave is not periodic.

\item for $\nu>0$, $\kappa=0$ or $\nu=0$, $\kappa>0$:  $\lambda_{1,2}=\alpha\pm \,  \i\,  \beta $, $\lambda_3\in \mathbb R$, the traveling wave is not periodic (this situation was considered in Sec.6).
\end{itemize}

The numerical computations show that the traveling waves for the nonlinear equation exist only on a bounded interval, the picture is very similar to Fig.1, right.

\section{8. QUASI-ONE-DIMENSIONAL MODELS OF BLOOD FLOW IN VESSELS}

A commonly used model of quasi-one-dimensional  blood flow in vessels was introduced in \cite{Brook1999}, see also \cite{Sherwin}, \cite{AmRo}.
In fact it is a gas-dynamics-like system of conservation laws
\begin{eqnarray*}\label{q1}
\dfrac{\partial S }{\partial t}  +  \dfrac{\partial S V}{\partial x} =0,\\
\dfrac{\partial SV }{\partial t} + \dfrac{\partial SV^2}{\partial x} +\frac{S}{\rho} \dfrac{\partial P}{\partial x}=F,\\
P(S) = P_0 + D(S - S_0),
\end{eqnarray*}
where
 $S(t,x)$ is the cross-sectional area of the vessel, $ V (t,x)$ is the velocity of blood flow,
$P(t,x)$ is the pressure, $F(t,x, S, V) $ is some external force,
 $\rho=\rm const$  is the density of blood, constants $P_0$, $S_0$ are the average pressure and average cross-sectional area,
 $D$ is the rigidity of the wall.

The main problem of this model is that for imitation of the blood pulsation it is necessary to add a periodical with respect to $t$ exterior force $F$. In other words, oscillations are imposed from outside, and are not determined by the properties of the system itself. This situation can be corrected by modeling the response of the vessel wall to the passage of a fluid flow. It turns out that spontaneous pulsations can be maintained if the walls of the vessel contract according to a certain law \cite{Ibatullin}.

Thus, we assume that $F=S E,$ where  $E=-\int\limits_{-\infty}^x \,(S-S_0)\, dx<\infty$.
The sense of this assumption is the following: the acceleration of velocity at a given point is proportional to the cross-sectional area of the vessel at that point and to the total volume of fluid that can flow through that point.

Then
\begin{eqnarray*}\label{BF}
\dfrac{\partial V }{\partial t}  + V \dfrac{\partial V}{\partial x}=- E + \mu \dfrac{\partial^2 E}{\partial x^2}, \qquad
\dfrac{\partial E }{\partial t} + V \dfrac{\partial E}{\partial x} = S_0 V,
\end{eqnarray*}
where $\mu=\frac{D}{\rho},$ $P=P_0-D E_x.$

If we set $S_0 =1$, then in the limit $D\to 0$ (a small rigidity) we get system  (\ref{me}), where $(V, U)=(V, E)$, therefore as a corollary of Theorem \ref{T3} we obtain a criterion for the singularity formation in a smooth solution.

However, our aim is to obtain conditions for a normal cardiac activity, in other words, conditions for existence of
traveling waves. For
$(V,E)=({\mathcal V}(\xi), {\mathcal E}(\xi)),$ $\xi= x-wt, $ we get
\begin{eqnarray*}\label{q1}
(-w + {\mathcal V})\,{\mathcal V}'= - {\mathcal E} + \mu {\mathcal E}'',\qquad
(-w + {\mathcal V})\,{\mathcal E}'=  S_0 {\mathcal V}.
\end{eqnarray*}
It can be transformed to
\begin{eqnarray}\label{EV}
{\mathcal E}'= - \frac{S_0 \mathcal V}{{\mathcal V}-w},\qquad
{\mathcal V}'=\frac{{\mathcal E}({\mathcal V}-w)^2}{({\mathcal V}-w)^3+\mu w S_0}.
\end{eqnarray}
System (\ref{EV}) on the phase plane  $({\mathcal E}, {\mathcal V})$ has the only equilibrium ${\mathcal E}= {\mathcal V}=0$.
If $w^2>\mu S_0$ it is a center, what means that periodic small perturbations exist.
If $w^2<\mu S_0$ it is   a saddle, what imply nonexistence of a periodic solution.

In contrast with the cases considered in Secs.6 and 7, system (\ref{EV}) has
the first integral
\begin{eqnarray*}\label{q1}
{\mathcal E}^2 + S_0 {\mathcal V}^2 -{w\mu S_0^2} \, \frac{2{\mathcal V}-w}{({\mathcal V}-w)^2} =\rm const.
\end{eqnarray*}
Fig.2 represents  phase curves on the plane $({\mathcal E}, {\mathcal V})$ depending on the proximity of the initial deviation to the origin.
\begin{figure}[h!]
\centerline{
\includegraphics[width=0.4\columnwidth]{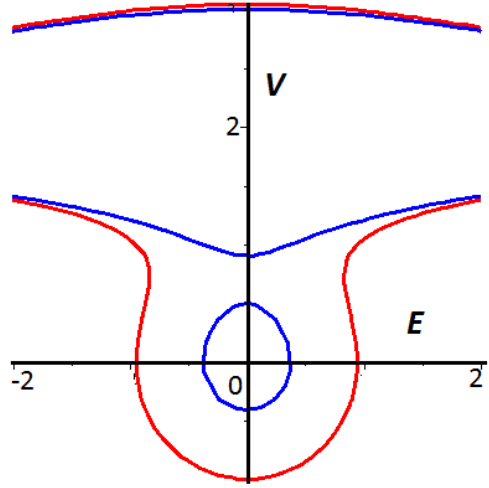}}
\caption{Periodic traveling wave for $w^2>\mu S_0$ ($\mu=S_0=1$, $w=2$).
Red: ${\mathcal E}(0)=0, {\mathcal V}(0)=-1$,  blue: ${\mathcal E}(0)=0, {\mathcal V}(0)=0.5$.
Periodic traveling wave exists for ${\mathcal E}(0)=0, 0<{\mathcal V}(0)<w-(\mu S_0 w)^{1/3}$, $w>0$.
}
\end{figure}

If a traveling wave exists, then we can close the vessel to obtain a simplest model of vicious circle of blood circulation. Let us compute a diameter $L$ of the closed vessel.

We denote
\begin{eqnarray*}\label{q1}
\Psi({\mathcal E}, {\mathcal V})={\mathcal E}^2 + S_0 {\mathcal V}^2 -{w\mu S_0^2} \, \frac{2{\mathcal V}-w}{({\mathcal V}-w)^2}.
\end{eqnarray*}
Assume that the initial data $({\mathcal E}(0), {\mathcal V}(0))$ are such that the motion is periodic. Let $V_-<0$ and $V_+>0$ be
the closest to zero roots of
\begin{eqnarray*}\label{q1}
S_0 {\mathcal V}^2 -{w\mu S_0^2} \, \frac{2{\mathcal V}-w}{({\mathcal V}-w)^2}=\Psi({\mathcal E}(0), {\mathcal V}(0)).
\end{eqnarray*}
Then the period of motion along the phase curve is
\begin{eqnarray*}\label{q1}
L&=&2 \int\limits_{V_-}^{V_+} \,\frac{({\mathcal V}-w)^2}{({\mathcal V}-w)^3+\mu w S_0}\,{\mathcal E}({\mathcal V})\, d{\mathcal V},\\
{\mathcal E}({\mathcal V})&=& \sqrt{\Psi({\mathcal E}(0), {\mathcal V}(0))- S_0 {\mathcal V}^2 +{w\mu S_0^2} \, \frac{2{\mathcal V}-w}{({\mathcal V}-w)^2}}.
\end{eqnarray*}

In addition, the opposite result can be obtained: for each closed vessel of perimeter $L$, there exists a constant $w$, corresponding to the velocity of the traveling wave inside this vessel.

\section{9. PRESSURELESS EULER-POISSON EQUATIONS}

The pressureless Euler-Poisson system is one of the most important models describing compressible media in the presence of a potential force, provided that the Laplacian of the potential depends on the density\cite{ELT}. It has the form
\begin{eqnarray}\label{EP}
\dfrac{\partial n }{\partial t} + \Div(n \bV)=0,\qquad
\dfrac{\partial \bV }{\partial t} + \left( \bV \cdot \nabla \right)
\bV =\,k \,  \nabla \Phi - q \bV  +\nu \Delta \bV,  \qquad  \Delta \Phi =n-n_0,
\end{eqnarray}
where  $n$ is the density,
$\Phi$ is the force potential, the vector $\bV$ is the velocity,
$n_0={\rm const}\ge 0$ is the density background,
$q={\rm const}\ge 0$ is the friction coefficient,
 $k=\rm const$,  the sign of which corresponds to the type of force. Namely, if
$k>0$, the  force is repulsive (it arises in models of plasma and semi-conductors),
if $k<0$, the  force is attractive (it arises in astrophysics).
If $k=0$,  system (\ref{EP}) decomposes and its part that relates to density and velocity corresponds to the pressureless gas dynamics.

System  (\ref{EP}) can be written in other terms. Namely, let us
introduce $\bE$ as
$
n=n_0- \Div \bE,$
 and remove $ n $.

The resulting system is
\begin{eqnarray*}\label{4}
\dfrac{\partial \bV }{\partial t} + \left( \bV \cdot \nabla \right)
\bV = \, - k \bE - q \bV + \Delta \bV,\qquad \frac{\partial \bE }{\partial t} + \bV \Div \bE
 = \bV.
\end{eqnarray*}
In  1D case it takes the form
\begin{eqnarray*}\label{EP1}
\dfrac{\partial V }{\partial t}  + V \dfrac{\partial V}{\partial x}=- k E -q V +\nu \dfrac{\partial^2 V}{\partial x^2},\qquad
\dfrac{\partial E }{\partial t} + V \dfrac{\partial E}{\partial x} = n_0 V.
\end{eqnarray*}
Let us dwell on the non-dissipative case $\nu=0$. Then
\begin{itemize}
\item for $k>0$, $n_0>0$, $q=0$ we get  the equations of a cold plasma ($k=n_0=1$), just considered in Sec.6;
\item for $k>0$, $n_0>0$, $q>0$ we get the equations of a frictional cold plasma \cite{RChD2020};
\item for $k<0$, $n_0=0$,

for $k>0$, $n_0>0$,

for $k>0$, $n_0=0$,

we also obtain non-strictly hyperbolic systems of equations, but their behavior differs from one of cold plasma equations
see \cite{ELT}.
\end{itemize}

In all these cases Theorem \ref{T3} can be applied for obtaining the criterion of the singularity formation for solutions to the Cauchy problem, the generalized solution can be defined after the moment of singularities formation and  simple and traveling waves can be constructed.

\section{10. DAVIDSON'S MODEL}

The influence of the magnetic field to the cold (collisional) plasma in the simplest case can be described by the model, introduced by
R. C. Davidson and P. P. Schram in \cite{Dav68} and became popular after the appearance of the book  \cite{Dav72} in 1972. It belongs to the type (\ref{msh}) and consists of three equations:
\begin{eqnarray*}
\dfrac{\partial V_1 }{\partial t} + V_1 \dfrac{\partial V_1}{\partial x} &=& -q V_1
 - E_1 - B_0\,V_2,
\\
\dfrac{\partial V_2 }{\partial t} + V_1 \dfrac{\partial V_2}{\partial x}&=& - q V_2 + B_0\, V_1,
\\
\dfrac{\partial E_1 }{\partial t} + V_1 \dfrac{\partial E_1}{\partial x} &=&  V_1.
\end{eqnarray*}
Here
 $
 \bV=(V_1(t,x), V_2(t,x), 0),\quad \bE(t,x) = (E_1(t,x),0,0),
$ where $ (x, y, z) $ are the Cartesian coordinate,
${\rm rot}\, \bE = 0$ (the electric field is irrotational),
the magnetic field does not depend on time and space,  directed along  $z$,
 $\bB(x,t) = (0,0,B_0)$, $B_0 \equiv {\rm const}$.
Davidson's model as a non-strictly hyperbolic system with
$Q= \left(\begin{array}{ccc}-q &-B_0 & -1\\ B_0 & -q &0\\1 & 0 & 0 \end{array}\right).$


For non-collisional case $q=0$ the following results were obtained in \cite{ChR2021D}:

\begin{itemize}
\item A criterion for a singularity formation:
\begin{theorem}
For the existence of a $ C^1 $ -- smooth $ \frac {2 \pi} {\sqrt {1 + B_0^2}} $ - periodic solution  $ (V_1,  V_2,  E) $ it should be
 \begin{eqnarray*}
\left( (V_1^0)' \right)^2 + 2 \, (E^0)' +2 B_0\, (V_2^0)' - B_0^2  -1 <0.
\end{eqnarray*}
at any point $ x \in \mathbb R $.
\end{theorem}
\item The increasing of $|B_0|$  basically leads to an expansion of the class of initial data providing the global smoothness.

\item The traveling waves exist (the explicit formula).
\end{itemize}

For collisional case $q>0$ the following results were obtained in \cite{DR2022}:

\begin{itemize}
\item A criterion for maintaining the global smoothness of the solution in terms of the initial data and the coefficients $ B_0 $ and $ q $.

\item  The initial data are divided into two classes: one leads to stabilization to the equilibrium, and the other leads to the destruction of the solution in a finite time.

\item
For small $ |B_0| $ an increase in the intensity factor first leads to a change in the oscillatory behavior of the solution to monotonic damping, which is then again replaced by oscillatory damping.

 \item
 At large values of $ |B_0| $, the solution is characterized by oscillatory damping regardless of the value of the intensity factor $ q $.

 \item  Both the presence of an external magnetic field of strength $ B_0 $ and a sufficiently large collisional factor $ q $ help to  suppress the formation of a finite-dimensional singularity.
\end{itemize}

\section{DISCUSSION}

This paper is a survey of methods that can be applied to nonstrictly hyperbolic systems of a particular type. Initially, such systems arose in the study of equations simulating cold plasma, but later it turned out that there are many other models of this type. We tried to mention all that we know today. For some of these models, in particular those describing cold plasmas, the behavior of smooth solutions (or solutions up to the moment of loss of smoothness) for the inviscid case has been studied quite well; we provide references to the corresponding works. However, the study of generalized solutions even for the simplest models is a completely open question. The same applies to the study of the method of vanishing viscosity, and in general the influence of the viscosity matrix on the behavior of the solution. Therefore, we can consider that the paper contains a program of further research, some parts of which may turn out to be very difficult.

\section{ACKNOWLEDGMENTS}
Supported by the Moscow Center for
Fundamental and Applied Mathematics under the agreement
N 075-15-2019-1621.

\end{document}